# Bacterial Foraging Optimized STATCOM for Stability Assessment in Power System

Shiba R. Paital[a], Prakash K. Ray[a], Asit Mohanty[b], Sandipan Patra[c], Harishchandra Dubey[d]

[a]Department of Electrical Engineering, IIIT, Bhubaneswar, India. (e-mail: shiba.paital@gmail.com; pkrayiiit@gmail.com),
[b]Department of Electrical Engineering, CET, Bhubaneswar, India. (e-mail: asithimansu@gmail.com), [c] School of Electrical and Electronic Engineering - DIT, Dublin (e-mail: patra.sandipan@gmail.com). [d]Department of Electrical Engineering, The University of Texas at Dallas, USA (e-mail: harishchandra.dubey@utdallas.edu)

*Abstract*— This paper presents a study of improvement in stability in a single machine connected to infinite bus (SMIB) power system by using static compensator (STATCOM). The gains of Proportional-Integral-Derivative (PID) controller in STATCOM are being optimized by heuristic technique based on Particle swarm optimization (PSO). Further, Bacterial Foraging Optimization (BFO) as an alternative heuristic method is also applied to select optimal gains of PID controller. The performance of STATCOM with the above soft-computing techniques are studied and compared with the conventional PID controller under various scenarios. The simulation results are accompanied with performance indices based quantitative analysis. The analysis clearly signifies the robustness of the new scheme in terms of stability and voltage regulation when compared with conventional PID.

## I. INTRODUCTION

Due to the deregulation in the energy market, complexity in configuration of power system network increases. The power plants are forced to operate near their maximum power transfer capability because of increased energy demand. Due to which system parameter and voltage profile changes suddenly from its specified value when subjected to disturbance. Reactive power compensation is a major problem in electrical networks which is also a major reason for voltage collapse. This may further result in complete blackouts [1]. Therefore reactive power is required to be controlled in order to reduce the difference between power generation and load demand. In order to improve the quality, stability and reliability of power supply. It can also lead to transient stability oscillations produced due to the shortage of damping torque in the rotor [2]-[4].

In the past decades, power system stabilizers (PSS) are used for decreasing the low frequency transient oscillations by supplying a modulating signal when the system is subjected to disturbance. But this technique of suppressing the low frequency oscillations fails under severe disturbances. Hence to overcome these demerits of the conventional PSS, new power electronics based technology called Flexible Alternating Current transmission system (FACTS) [4] is evolved. The primary objectives of these controllers are to regulate the power flow, improve the system stability, power transfer capability and to suppress power system oscillations produced due to disturbance. FACTS devices like static var compensator (SVC), STACOM, unified power flow controller (UPFC) etc. [5]-[13] which can either be connected in series, shunt or in series-shunt manner and operates primarily depending on their location and type of connection. Among these FACTS devices, one of the most promising shunt connected FACTS controller is called STATic COMpensator (STATCOM) which was considered in this paper for reactive power compensation [8]-[9], regulation of bus voltage, improvement of dynamic behavior of the system as compared to the conventional SVC. It uses components such as coupling transformer, DC capacitor with a six pulse DC to AC voltage source inverter (VSI) for converting DC to 3-φ AC voltage with constant frequency, amplitude and phase .STATCOM has possibly two steady state modes of operation namely inductive (lagging) and Capacitive (leading) so that it can supply or consume reactive power to maintain the bus voltage without using any reactors or any large capacitor banks. Gate turn-off (GTO) based STATCOM damping stabilizer is considered. So as to act as an alternative to the conventional SVC as it provides better dynamic and transient stability characteristics [10]-[12]. In this study, some heuristic techniques such as PSO, BFO are used for tuning the parameters of PID controller. It can be clearly understood from the simulation results that PSO-PID, BFO-PID gives better responses in terms of stability as compared to other techniques.

## II. SYSTEM UNDER STUDY

In this section, SMIB with a STATCOM controller is described for the stability study. The SMIB is incorporated with PSS along with a STATCOM to improve the transient stability. This system is shown in Fig. 1. Here the voltage profile and low frequency oscillations are improvised by the quick response of the proposed controller. The configuration and its Linearized model is shown in Fig. 1 (a) and (b) respectively.

Fig.1(a) comprises of a three phase gate turn-off (GTO) based voltage source inverter (VSC) and a DC capacitor based STATCOM connected to a SMIB system through a step-up transformer, where $Z_1 = R_1 + jX_1$, $Z_2 = R_2 + jX_2$, $Y_L = g + jb$, $x_t$ represents the line impedances and load admittance, the leakage reactance of transformer respectively. Fig.1 (b) represents the linearized block diagram of a SMIB system with STATCOM. Linearized model was considered in order to study the low frequency transient stability oscillations with a smaller variation of load [9], [11].

The output of the voltage source converter (VSC) is given by,

$$v_s = CV_{dc}\angle\phi \tag{1}$$

Where, $v_s$ represents controllable ac voltage, $C = mk$, $m$ is the modulation ratio defined by pulse width modulation (PWM), $k$ is a constant defined as ac to dc voltage, $V_{dc}$ the dc-link voltage, and $\phi$ is the phase angle in PWM. The magnitude and the phase of the voltage $v_s$ is regulated by $m$ and $\phi$ respectively and the dc voltage is given by,

$$\dot{V}_{dc} = \frac{I_{dc}}{C_{dc}} = \frac{I_{dc}}{C}\left(i_{sd}\cos\phi + i_{sq}\sin\phi\right) \tag{2}$$

Where, $I_{dc}$ represents the current through the capacitor, $i_{sd}$ and $i_{sq}$ represents the d-axis and q-axis component of $i_s$ respectevely. Fig.2. represents the control structure of STATCOM with PID controller and lead lag damping stabilizer. $V_m$ and $V_m^{ref}$ are considered as the input to the controller, $K_P$, $K_I$, $K_D$ are the PID controller gains. Where $\Delta\omega$ is considered as the input to the lead lag damping controller.

(a)

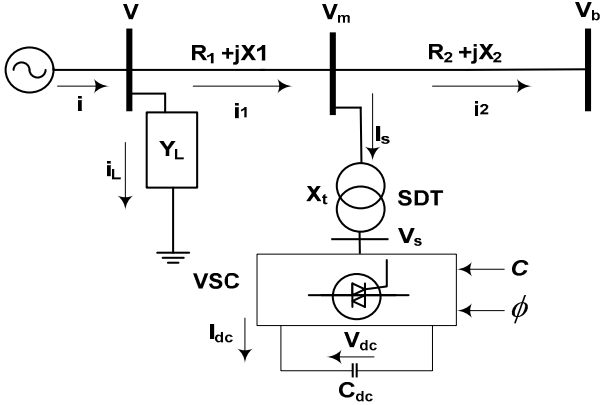

(b)

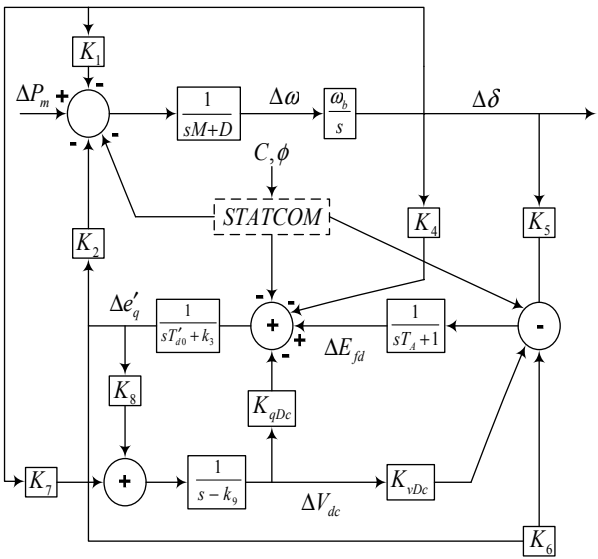

Fig. 1. Single machine infinite bus system with a STATCOM (a) configuration (b) linear model.

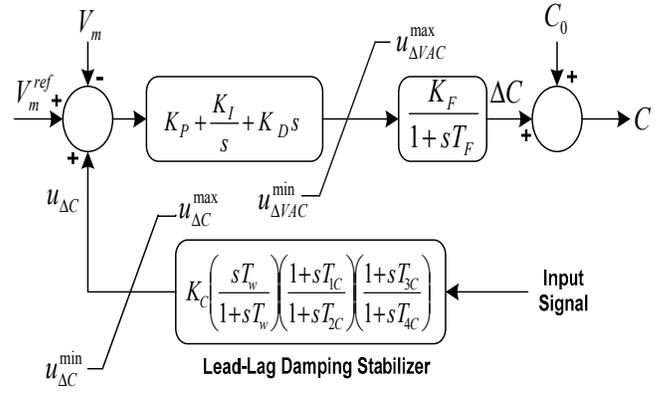

Fig.2. STATCOM with PID controller for ac voltage with a lead–lag damping stabilizer.

### III. PROBLEM FORMULATION

#### A. Stabilizer structure

The control structure of STATCOM based controller shown in Fig.2. consists of a lead-lag damping stabilizer for stability study. The transfer function of the lead-lag damping stabilizer is given by:

$$u = K_C\left(\frac{sT_W}{1+sT_W}\right)\left(\frac{1+sT_{1C}}{1+sT_{2C}}\right)\left(\frac{1+sT_{3C}}{1+sT_{4C}}\right)\Delta\omega \tag{3}$$

Where $K_C$ is the stabilizer gain, $T_W$ is the washout time constant. $T_{1C}$ and $T_{2C}$ are the lead time constants, $T_{3C}$ and $T_{4C}$ are the lag time constants, change in speed $\Delta\omega$ is the input to the stabilizer and $u$ is the output of the stabilizer [1]-[2]. Washout and lag time constants are usually predefined whereas controller gain and lead time constants are to be determined.

#### B. Objective function

The performance indices of the purposed controller was investigated in order to minimize the electromechanical oscillations in terms of Integral of the Time-Weighted Absolute Error (ITAE) with PID controller to minimize the change in ac and dc voltage for STATCOM in terms of their time-domain responses in order to improve the power system stability of the system. Here a multi-objective function is being formed with a coordination of damping and PID controller.

$$ITAE = J = \int_{t=0}^{t=t_{sim}} t(|\Delta\omega| + \gamma_1|\Delta V_m| + \gamma_2|\Delta V_{dc}|)dt \tag{4}$$

Where, $t_{sim}$, $\Delta\omega$, $\Delta V_m$, $\Delta V_{dc}$, $\gamma_1$, $\gamma_2$ represents the simulation time, change in speed, change in dc bus voltage and weighting factors respectively. This paper aims to optimize the objective function through different optimizing techniques like PSO and BFO in order to improve the time domain specifications and low frequency oscillations respectively.

## C. Conventional Proportional Integral Derivative (PID) Controller

In the current study for stability analysis, a PID is used as a conventional controller for reactive power compensation. The basic diagram to represent PID controller is shown in Fig. 3. The transfer function of the controller in frequency and time domain is given by;

$$TF_{PID}(s) = K_p + \frac{K_i}{s} + K_d s \quad (5)$$

$$u(t) = K_p e(t) + K_i \int_0^t e(t)dt + K_d \frac{d}{dt}e(t) \quad (6)$$

Where u(t) is the control signal and e(t) is the error signal.

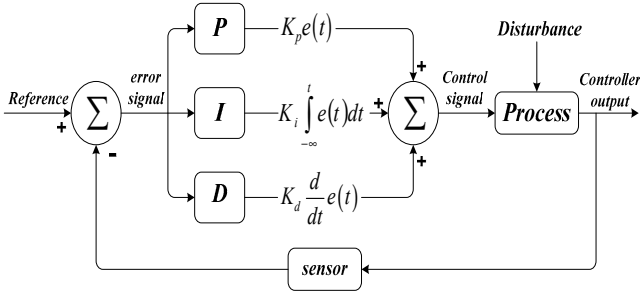

Fig.3. Schematic diagram representing PID controller.

## D. PSO-PID Controller

Particle Swarm Optimization (PSO) is one of the heuristic algorithm proposed by Kennedy and Eberhaut in 1995, which is based on the behaviour of the swarm in search of their food in a multi dimensional space based on their own and group members experience such as in Fish schooling, bird flocking practice etc. [14]-[15] PSO due to its high efficiency quick and simplest structure it is very much popular in the past years for tuning of PID controllers. For tuning of PID controllers or getting the optimal values ($K_p, K_i$ & $K_d$), each particle in the entire swarm search to update their local best position ($p_{best}$) and velocity to global position ($g_{best}$) and velocity as shown in Fig. 4.

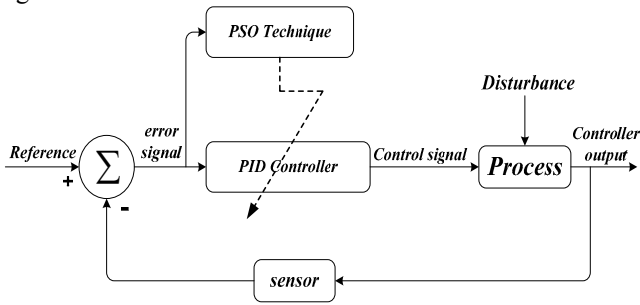

Fig. 4. Block diagram of proposed PSO-PID technique.

The algorithm of PSO can be described in steps as follows:
Step 1: Initialize the particle and population size, dimension of the search space with random position and velocity.
Step 2: Calculate desired fitness function of each particle.
Step 3: Comparison of the found fitness of each particle to that of its current fitness value. If the previous fitness value is better, update present fitness value and update its best position.
Step 4: Comparison of fitness of each particle to that of global best fitness value of the swarm. If the current value is better, replace the global best fitness and update global best position.
Step 5: Updation of particle's position and velocity by using (7) and (8).
Step 6: Repeat step 2-5 till the converging criterion is satisfied.

The velocity and position of each particle can be updated using the following equations:

$$v_j^d(t+1) = w(t) * v_j^d(t) + C_1 * r_1 \left( pbest_j^d - x_j^d(t) \right) \quad (7)$$
$$+ C_2 * r_2 \left( gbest_j^d - x_j^d(t) \right)$$

$$gbest_j^d = x_j^d + v_j^d(t+1) \quad (8)$$

Where $v_j^d(t)$ is the jth particle in the dth dimension at iteration t, $x_j^d(t)$ is the current position of particle, $C_1$ and $C_2$ are acceleration coefficients, $r_1$ and $r_2$ represents random numbers within [0,1].

## E. BFO-PID Controller

Bacteria foraging optimization proposed by Passino adopts the foraging strategy of Escherichia coli (E.coli) bacteria for solving the real world optimization problems. Foraging is the process of locating food by a group of E.coli bacterium found in the human intestine and this bacterium forages through locating, handling and ingesting food. In this process of foraging, every bacterium remains in the search for the food with avoiding the noxious substances and always tries try to make the most of it. In this process of foraging, Bacterium's Communicates through sending signals with each other. This optimization process mainly rely on four sequential mechanisms namely chemotaxis, swarming and reproduction and elimination-dispersal [16].

The algorithm of Bacteria foraging optimization can be summarized as follows:
Step 1: Initialize all the variables i.e. Population size N, Search space dimension D, stopping criteria ($J_{min}$), Proportional gain ($K_p$), integral gain ($K_i$), derivative gain ($K_d$) etc.
Step 2: Evaluate the objective function for the optimization process.
Step 3: (i) Start of the swimming or tumbling operation, (ii). Start of the elimination-dispersal loop, (iii). Start the reproduction loop, (iv). Start the chemotaxis loop.
Step 4: Update the fitness of each bacteria based on their health.
Step 5: If the stopping criteria are met, stop the search and display the parameter values. Otherwise, repeat step 1 to step 4 until the optimal values are obtained.

## IV. SIMULATED RESULTS AND DISCUSSIONS

This section describes the simulation results with their performance descriptions. The single machine connected to infinite bus is being modelled in MATLAB/Simulink in the form of linearized transfer functions. The basic assumption of small disturbances in load in the form of variation in real and reactive power is main drive of designing the linear model of the system. Here, it may be urged that though a linearization of the non-linear may affect the system response or performance for large variation in system parameters. But, it will be less affected when the small-signal stability analysis with a smaller variation in parameters are coming to the system as disturbances.

*A. Stability test under normal loading*

For verification of the performance of the proposed stabilizers under normal loading conditions, the following variations are considered which is caused by a fault disturbance. The proposed controllers are tested with the following loading conditions:

(a) nominal loading (P, Q, V) = (0.7, 0.3, 1.0) pu;

It is observed that when the SMIB system is simulated with STATCOM controller with PID controller, PSO-PID and BFO-PID, the variation in system voltage is providing a control in oscillation around the nominal voltage of 1 pu as shown in Fig. 5 (a). The corresponding variation in system angular frequency is shown in Fig. 5 (b). It is clearly noticed that after a lot of hit-and-trial PID controller is still giving some oscillations, whereas both PSO and BFO are stabilizing the disturbance in better way with decreased settling times and peak overshoots.

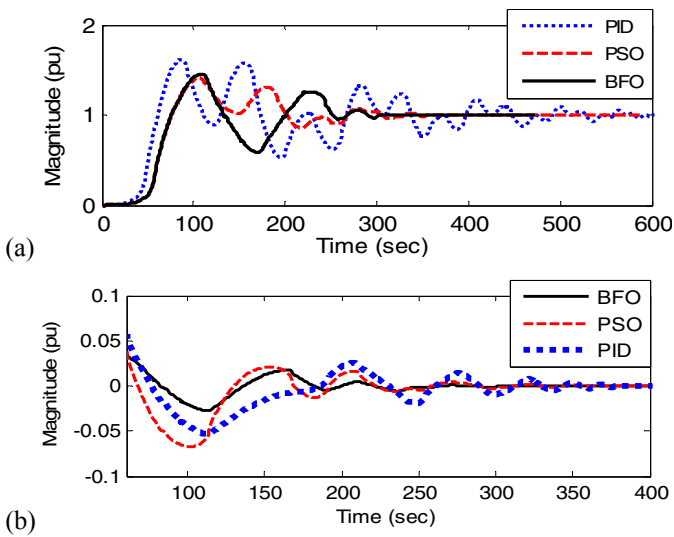

Fig. 5 Stability performance study of (a) voltage and (b) angular frequency in SMIB

*B. Stability test under smaller variation in loading*

Again, the stability performance of STATCOM in SMIB is tested under smaller variation loading conditions (10-20% w.r.t. the rated value). The variations considered are described below in terms of parameters caused by a fault. The variation in loading conditions is given by:

(b) nominal loading (P, Q, V) = (1.2, 0.4, 1.15) pu;

It is observed STATCOM with PID controller, PSO-PID and BFO-PID controller stabilizes the voltage of the SMIB in a way as shown in Fig. 6 (a) where the nominal voltage is 1 pu. Also, the variation in angular frequency under the same operating condition is shown in Fig. 6 (b). It is again observed that both PSO and BFO are stabilizing the voltage and frequency fairly better than the conventional PID controller designed on heuristic hit-and-trial tuning method and thus show decreased settling times and peak overshoots.

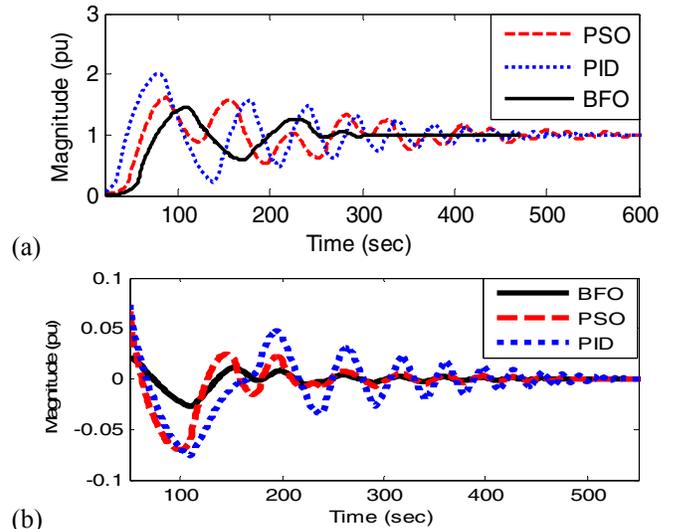

Fig. 6 Stability performance study of (a) voltage and (b) angular frequency in SMIB.

*C. Performance of BFO technique*

The gains of PID controller are being optimized by PSO and BFO. While performing the different steps of these algorithms, Accumulative fitness value and the probability of each step occurrence plays a very important role in obtaining the optimal values of $K_p$, $K_i$ & $K_d$. The performance of BFO in terms of these parameters during the whole iteration is shown in Fig. 7 (a, b) accumulative fitness and (c, d) probability. It shows that these values go on changing the iteration and finally the PID gains are chosen from the optimum point in the curves.

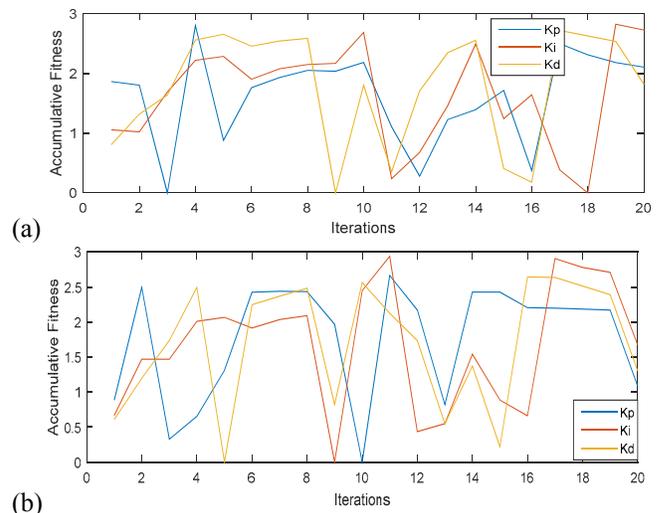

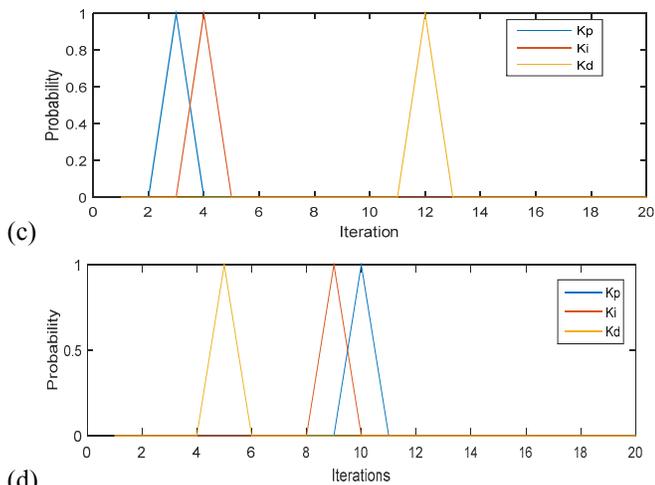

Fig. 7 Performance of BFO in terms of (a, b) Accumulative fitness and (c, d) probability

## V. CONCLUSION

In this paper, study of power system stability is improved by the use of STATCOM-based stabilizers under different operating conditions. The PID controller used in STATCOM is optimized by PSO as well as BFO to improve the system performance in terms stability. It is observed that the PSO and BFO optimized STATCOM controller provide a better transient performance as compared to the conventional PID controller. Again a comparative performance analysis is presented in terms of the performance indices; settling time and peak overshoot for different operating conditions which validates the improved performance of the proposed controller with respect to the conventional PID controller.